\documentclass{ws-procs9x6}

\newcommand{\be}{\begin{equation}}\newcommand{\ee}{\end{equation}}
\newcommand{\bea}{\begin{eqnarray}}
\newcommand{\eea}{\end{eqnarray}}
\newcommand{\beq}{\begin{eqnarray}}
\newcommand{\eeq}{\end{eqnarray}}
\newcommand{\al}{\alpha}

\def\({\left(}
\def\){\right)}
\def\[{\left[}
\def\]{\right]}

\begin{document}

%\title{Casimir interaction between a perfect
%conductor and graphene described by the Dirac model}

%\title{On Casimir force between ideal conductor
%and suspended graphene modeled with relativistic fermions}

\title{Suspended graphene films and their Casimir interaction 
with ideal conductor}

\author{I.~V.~Fialkovsky
\footnote{The author gladly acknowledge the financial support of FAPESP,
as well as of grants RNP 2.1.1/1575 and RFBR $07$--$01$--$00692$.}}
\address{Instituto
de F\'isica, Universidade de S\~ao Paulo,
S\~ao Paulo, S.P., Brazil\\
Department of Theoretical Physics, Saint-Petersburg
State University, Russia}

\begin{abstract}
We adopt the Dirac model for graphene and calculate the Casimir
interaction energy  between a plane suspended graphene sample and
a parallel plane ideal conductor. We employ both the Quantum Field Theory (QFT) approach,
and the Lifshitz formula generalizations. 
The first approach turns out to be the leading order in
the coupling constant of the second one. The Casimir interaction
for this system appears to be rather weak but experimentally measurable.
It exhibits a strong dependence on the mass of the quasi-particles in graphene.

Present article is based on joint works \cite{IFDV,grcas}.
\end{abstract}
%\pacs{12.20.Ds, 73.22.-f}
\bodymatter
%\vfill\eject
\section{Introduction}
Graphene is a (quasi) two dimensional hexagonal lattice of carbon
atoms. It belongs to the most interesting materials in
solid state physics now due to its exceptional properties and
importance for nano technology \cite{gra-rev,RMP}. Here
we consider the Casimir interaction between suspended graphene
plane and parallel ideal conductor. This setup was considered in
\cite{BV,Bordag:2005by,BGKM} using a hydrodynamical model for the
electrons in graphene following \cite{Fetter73,BIII}. Later it
became clear that this model does not describe the electronic
properties specific to this novel material.

Here we use a
realistic and well-tested model where the quasi-particles in
graphene are considered to be two-component Dirac fermions.
This model incorporates the
most essential and well-established
properties of the
their %quasi-particles'
dynamics: the symmetries of the hexagonal lattice,
 the linearity of the spectrum, a very small mass gap
(if any), and a characteristic propagation velocity which is $1/300$
of the speed of light \cite{gra-rev,DiracModel}.
By
construction, this model should work below the energy scale of
about $1eV$, but even above this limit the optical properties
 of graphene are reproduced with a high precision
\cite{Nair}.

The action of the model, therefore,
is given by
\begin{equation}
S_{\rm D}=%\sum_N
    \int d^3\,x \bar\psi (\tilde \gamma^l
    (i\partial_l-eA_l)-m)\psi, \label{Di}\quad l=0,1,2
\end{equation}
where $\tilde\gamma^l$ are just rescaled  $2^\times2$ gamma matrices,
$\tilde\gamma^0\equiv\gamma^0$, $\tilde\gamma^{1,2}\equiv
v_F\gamma^{1,2}$, $\gamma_0^2=-(\gamma^i)^2=1$.
In our units, $\hbar=c=1$, and Fermi velocity  $v_F\simeq
(300)^{-1}$.
%In graphene the overall number $N$ of two-component
%fermions is equal to four due to symmetries of the carbon lattice.
The value of the mass gap parameter $m$ and
mechanisms of its generation are under discussion
\cite{Appelquist:1986fd,massgap1,Gusynin,Pyatkovskiy}. The upper
limit on $m$ is about $0.1eV$ at most.

The propagation of photons in the ambient $3+1$
dimensional space is described by the Maxwell action
\begin{equation}
S_{\rm M}=-\frac 14 \int d^4x F_{\mu\nu}F^{\mu\nu},\quad
\mu,\nu=0,1,2,3. \label{SEM}
\end{equation}
%The coupling constant is normalized according to
%$e^2/(4\pi)=\alpha\simeq 1/137$.
%
In the following we shall suppose that the graphene sample occupies the
plane $x^3=a>0$, and the conductor corresponds to $x^3=0$.\\

%\vfill\eject
%%%%%%
\section{QFT approach}
In the framework of QFT one evaluates the effective action in a
theory described by the classical action $S_{\rm D}+S_{\rm M}$.
Then the Casimir energy density per unit area of the
surfaces at the leading order in the fine
structure constant $\alpha$  is given by
\begin{equation}
\mathcal{E}_1= -\frac 1{TS}
\raisebox{-3.75mm}{\psfig{figure=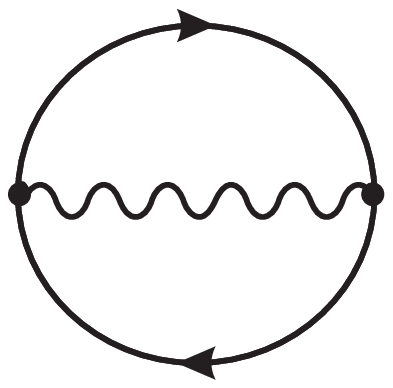,height=.4in}},
\label{E1}
\end{equation}
where  $T$ is time interval, and  $S$ is the
area of the surface.
The solid line denotes the fermion propagator in $2+1$
dimensions (i.e., inside the graphene sample), and the wavy line
is the photon propagator in the ambient $3+1$ dimensional space
subject to the perfect conductor boundary conditions at $x^3=0$:
$
A_0\vert_{x^3=0}=A_1\vert_{x^3=0}=A_2\vert_{x^3=0}=\partial_3A_3\vert_{x^3=0}=0
%\label{condbc}
$.

The fermion loop in $2+1$ dimensions has already been
calculated in a number of papers
\cite{Appelquist:1986fd,Gusynin,Pyatkovskiy}.
It gives the quadratic order in $A$ of the effective action
for electromagnetic field
%\begin{eqnarray}
\be
S_{\rm eff}(A)= A \ \raisebox{-3.75mm}{\psfig{figure=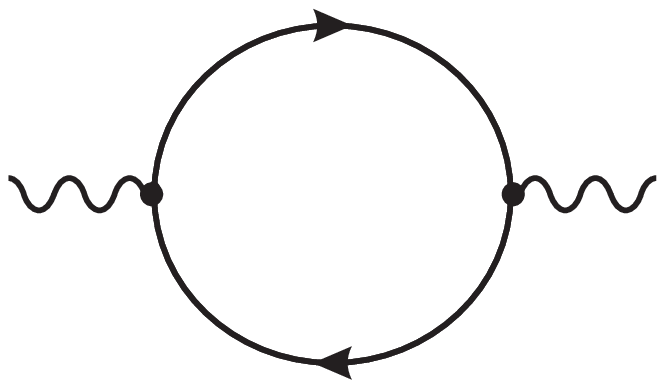,height=.4in}}\  A
= \frac 12 \int \frac{d^3p}{(2\pi)^3} A_j(p)
\Pi^{jl}(p)A_l(p),\label{Seff}
\ee
%\end{eqnarray}
where
\begin{eqnarray}
 &&\Pi^{mn}
        = \frac {\alpha\Phi(\tilde p)}{v_F^2} \,
             \eta^{m}_{j}\left(g^{jl}-\frac{\tilde p^j\tilde p^l}{\tilde p^2}\right)\eta_l^n\,,
\label{Pmn}
\end{eqnarray}
is the polarization tensor in the lowest, one loop, order.
Here $\eta^{m}_j={\rm diag}(1,v_F,v_F)$,  $\tilde p$ denotes 
the rescaled momenta $\tilde p_j=\eta_j^k p_k$, 
The function $\Phi({p})$ is model dependent, and for graphene it reads
%and for graphene 
$\Phi({p})=
        2\(2 m \tilde p -(\tilde p^2+4m^2){\rm arctanh }({\tilde p}/{2m})\)/{\tilde
        p}$.
We assume here that all possible parity-odd parts are canceled in $\Pi$.
%if not see \cite{IFDV}, 
Possible effects invoked by their presence are considered in \cite{IFDV}. 
%for rough numerics see also \cite{MHA}.

To calculate the diagram (\ref{E1}) we only need to couple the
kernel (\ref{Pmn}) to the photon propagator subject to conducting boundary conditions.
%and integrate over the
%photon momenta.
In Fourier representation
and for the Euclidean $3$-momenta, i.e., after the Wick rotation
$p\to p_E=(ip_0,p_1,p_2)$,
%$p_4=$,
%the explicit form for
the $a$-dependent part of the energy
reads
\be%gin{eqnarray}
\mathcal{E}_1
   =-\frac 14 \int\frac{d^3p_E}{(2\pi)^3}\,
\frac{\Pi_j^j(p_E)}{p_\|}\, e^{-2ap_\|} % \nonumber\\
=-\int\frac{d^3p_E}{(2\pi)^3}\,
        \frac{\al (p_\|^2+\tilde p_\|^2)\Phi(p_E)}{4p_\|^{\phantom{2}} \tilde p_\|^2}\, e^{-2ap_\|}\,. \label{E11}
\ee%nd{eqnarray}
where we expanded  $\Pi_j^j(p_E)$ explicitly  with help of
(\ref{Pmn}), and $p_\| \equiv |p_E|$.
%For further details of the calculations see \cite{grcas}.

%\vfill\eject
%%%%%%%%%%%%%%%%%%%%%%%%%%%%%%%%%%%%%%%%%%%%%%%%%%%%%%%%%%%%%%%%%%%%%%%%%%%
%%%%%%%%%%%%%%%%%%%%%%%%%%%%%%%%%%%%%%%%%%%%%%%%%%%%%%%%%%%%%%%%%%%%%%%%%%%%%%%%%%%%%%%%%%%%%
\section{Lifshitz formula approach}
One can also consider the system as described by effective theory of the
electromagnetic field with the action $S_{\rm M}+S_{\rm
eff}$ subject to the conducting boundary conditions %(\ref{condbc})
at $x^3=0$. Then at the
surface of graphene, the Maxwell equations receive a singular
contribution
\begin{equation}
\partial_\mu F^{\mu\nu}+\delta(x^3-a)\Pi^{\nu\rho}A_\rho =0\label{Meq}
\end{equation}
following from $S_{\rm eff}$. Here we set $\Pi^{3\mu}=\Pi^{\mu3}=0$. This %singular
contribution is equivalent to imposing the matching conditions
\be
 (\partial_3 A_\mu)\vert_{x^3=a+0}-(\partial_3A_\mu)\vert_{x^3=a-0}
    = \Pi_\mu^{\ \nu}A_\nu\vert_{x^3=a} \,.\label{match}
\ee
assuming that $A_\mu$ is continuous at $x^3=a$.
%At this stage,
Now, one can forget the origin of $\Pi_\mu^{\ \nu}$ and
quantize (at least formally) the electromagnetic field subject to the conditions
(\ref{match}) at $x^3=a$ and to the conducting conditions at
$x^3=0$.

The original Lifshitz approach \cite{Lifshitz}
was generalized \cite{Bordag:1995jz,Reynaud} for the
interactions between two plane parallel interfaces separated by the
distance $a$ and possessing arbitrary reflection coefficients
$r^{(1)}_{\rm TE, TM}$, $r^{(2)}_{\rm TE, TM}$ of the TE and TM
electromagnetic modes on each of the surfaces
\begin{equation}
    {\mathcal{E}}_L
    =\int\frac{d^3p_E}{16\pi^3} \ln [(1-e^{-2p_\| a}r_{\rm TE}^{(1)}r_{\rm TE}^{(2)})
        (1-e^{-2p_\| a}r_{\rm TM}^{(1)}r_{\rm TM}^{(2)})] .
        \label{EL}
\end{equation}

 For graphene with help of matching conditions
(\ref{match}) we can obtain at the Euclidean momenta
\be
    r_{\rm TE}^{(1)}=\frac {-\alpha \Phi}{2p_\| +\alpha\Phi}\,,\quad
    r_{\rm TM}^{(1)}=\frac {\alpha p_\| \Phi}{2\tilde p_\|^2 + \alpha p_\| \Phi}
    \label{rTETM},
\ee while for the perfect conductor one has
$r_{\rm TE}^{(2)}=-1$, $ r_{\rm TM}^{(2)}=1$.
It is clear, that $\Phi$ must be rotated to Euclidean momenta
as well. We also note that the perfect conductor case is recovered
from (\ref{rTETM}) in the formal limit $\Phi\to\infty$.

One can show by a direct computation that the energy
$\mathcal{E}_1$, Eq.\ (\ref{E11}), coincides with the leading
$\alpha^1$ order in a perturbative expansion of the Lifshitz
formula (\ref{EL})-(\ref{rTETM}), so that the two approaches are
consistent.
%In fact, the Lifshitz formula is the one-loop vacuum
%energy (one closed vacuum loop) in an effective theory
%corresponding to the action $S_{\rm M}+S_{\rm eff}$.

%%%%%%%%%%%%%%%%%%%%%%%%%%%%%%%%%%%%%%%%%%%%%%%%%%%%%%%%%%%%%%%%%%%%%%%%%%%%%%%%%%%
\section{Results and discussion}
The formulae (\ref{E11}) and (\ref{EL})-(\ref{rTETM}) are
suitable for the numerical and asymptotical evaluation.
The asymptotic expansion  for short and long distances  are readily obtained
through uniform expansion of the integrand of (\ref{E11}), (\ref{EL})
\be\label{EL2-3}
{\mathcal{E}}_L\raisebox{-5pt}
    {$\sim\atop a\to \infty$}-\frac{\al}{24\pi^2}\,\frac{2+v_f^2}{m a^4},
\qquad\quad
{\mathcal{E}}_L\raisebox{-5pt}
    {$\sim\atop a\to 0$}    \frac{1}{16 \pi a^3}\,h(\al,v_F)
\ee
Note that the asymptotics at large separations  is of the first order in
$\alpha$
while for small separations, it contains all powers of $\al$ through $h(\al,v_F)$, for
the real values of parameters in graphene $h(1/137,1/300)\approx 0.024$.

Therefore we see that at large separations
Casimir energy does not turn into the ideal conductor case,
while at small separation this case is indeed recovered. This is
counter-intuitive since the main contribution at short separations
shall come from the high momenta for which one would expect the
graphene film to become transparent.
We must also stress that this behavior is
drastically  different from that in the hydrodynamic model.
\cite{BV}-\cite{BGKM}

\begin{figure}%\label{fig1}
\begin{center}
\includegraphics[width=3.1in]{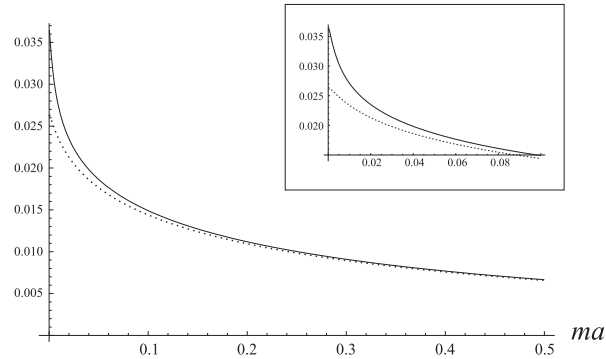}
\end{center}\caption{\label{fig1}The relative Casimir energy densities
$\mathcal{E}_1/\mathcal{E}_C$ (solid line) and
$\mathcal{E}_L/\mathcal{E}_C$ (dashed line) as functions of $ma$.
Insert shows a zoom of the small-distances region.}
\end{figure}

For numerical evaluation we normalize the results
to the Casimir energy
$%\begin{equation}
\mathcal{E}_C=-\frac{\pi^2}{720\, a^3} \label{ECas}
$ %\end{equation}
for two plane ideal conductors separated by the same distance $a$.
The results of calculations are depicted at Fig.\
\ref{fig1}.
The scale is defined by the mass parameter $m$.
For  $m$ of the order of next nearest-neighbor hopping energy $t'$,
i.e., $m= 0.1eV$ \cite{RMP}, $ma=1$ corresponds to $a=1.97$
micrometer.

Thus, we can see that the magnitude of the considered Casimir
interaction of graphene with a perfect conductor is rather small.
Actual measurement of such weak forces is a challenging, but by no
means hopeless, experimental problem. Strong dependence on the
mass parameter $m$ at large separation is also a characteristic
feature of the Casimir force. Getting an independent measurement
of $m$ may be very important for our understanding of the
electronic properties of graphene. The mass of quasi-particles in
graphene is, probably, very tiny, which improves the detectability
of the Casimir interaction since the energy increases with
decreasing $m$.

%\section*{Acknowledgments}

\bibliographystyle{ws-procs9x6}
%\bibliography{ws-pro-sample}

\end{document}